\documentclass[%
 reprint,
 nofootinbib,
 amsmath,amssymb,
 aps,
 prd,
 floatfix,
]{revtex4-2}

\usepackage{aas_macros}
\usepackage{graphicx}
\usepackage[dvipsnames]{xcolor}

\usepackage[colorlinks]{hyperref}
\hypersetup{
    colorlinks=true,
    linkcolor=blue,
    filecolor=blue,      
    urlcolor=blue,
    citecolor=magenta,
    pdfpagemode=FullScreen
}

\newcommand{\orcid}[1]{\hspace{1mm}\href{https://orcid.org/#1}{\includegraphics[height=0.3cm,keepaspectratio]{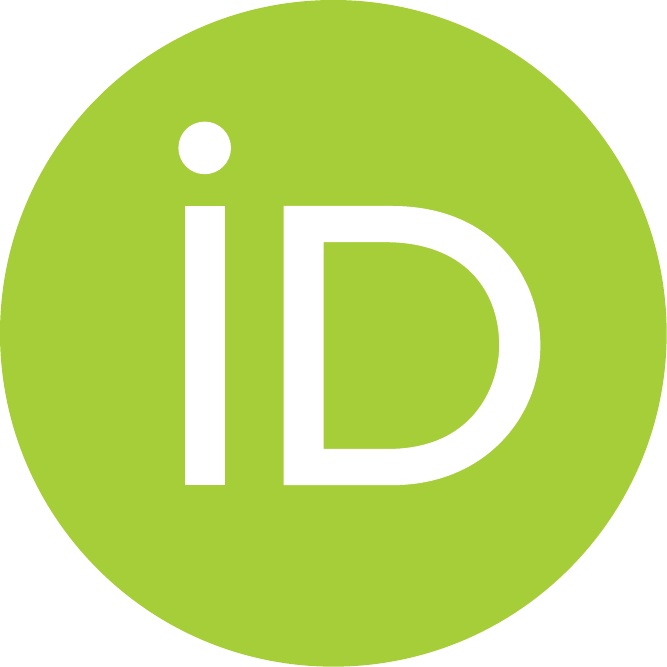}}}

\begin{document}

\title{Intermediate mass and heavy Galactic cosmic-ray nuclei: the case of new AMS-02 measurements}

\author{Benedikt Schroer\orcid{0000-0002-4273-9896}}
\email{benedikt.schroer@gssi.it}
\affiliation{Gran Sasso Science Institute (GSSI), Viale Francesco Crispi 7, 67100 L'Aquila, Italy}
\affiliation{INFN-Laboratori Nazionali del Gran Sasso (LNGS),  via G. Acitelli 22, 67100 Assergi (AQ), Italy}

\author{Carmelo Evoli\orcid{0000-0002-6023-5253}} 
\email{carmelo.evoli@gssi.it}
\affiliation{Gran Sasso Science Institute (GSSI), Viale Francesco Crispi 7, 67100 L'Aquila, Italy}
\affiliation{INFN-Laboratori Nazionali del Gran Sasso (LNGS),  via G. Acitelli 22, 67100 Assergi (AQ), Italy}

\author{Pasquale Blasi\orcid{0000-0003-2480-599X}}
\email{pasquale.blasi@gssi.it}
\affiliation{Gran Sasso Science Institute (GSSI), Viale Francesco Crispi 7, 67100 L'Aquila, Italy}
\affiliation{INFN-Laboratori Nazionali del Gran Sasso (LNGS),  via G. Acitelli 22, 67100 Assergi (AQ), Italy}

\date{\today}

\begin{abstract}
The recent measurement of the spectra of intermediate mass nuclei and iron nuclei carried out with the AMS-02 experiment provided us with the most complete set of data on cosmic ray fluxes to date, and allowed us to test the standard model for the transport of these particles through the Galaxy to the finest details. We show that the parameters derived from lighter primary and secondary elements in the cosmic radiation also lead to a good description of the data on heavier nuclei, with no need to invoke different injection spectra for such nuclei, provided the whole chain of fragmentation is properly accounted for. The only exception to this finding is represented by iron nuclei, which show a very unusual trend at rigidity $\lesssim 100$ GV. This trend reflects in a Fe/O ratio that is at odds with the results of the standard model of cosmic ray transport, and is in contradiction with data collected by HEAO, ACE-CRIS and Voyager at lower energy. We speculate on possible origins of such findings. 
\end{abstract}

\maketitle
\section{Introduction}

The transport of cosmic rays (CRs) in the Galaxy is a crucial piece in the puzzle that leads to understanding the origin of these high energy particles. The description of CR transport is based on a macroscopic description of the particle motion (advection and diffusion) that is based on microscopic processes (particle wave resonances and CR self-generation) and particle physics (radiative losses and nuclear fragmentation). As such, it represents an endeavour of remarkable proportions, that at best can be pursued by using effective theories. The advection-diffusion equation engulfs such difficulties in a way that led to numerous successes and that requires continuous adjustments to reflect pieces of observations that are being acquired. 

The precision measurements made by AMS-02 in the last decade or so have affected our understanding of the origin of CRs in a profound way, by confirming some pillar ideas and by revealing anomalies that might be signs of a need for a deep revision of some aspects of the problem: 1) the observation of spectral breaks in both primary and secondary nuclei ~\cite{PAMELA.2011.phe,AMS02.2015.protons,AMS02.2015.he,AMS02.2018.libeb} has driven us towards developing theories that could accommodate such features. The fact that the breaks appear also in the ratio of secondary-to-primary nuclei is a clear indication that they are most probably due to some new phenomenon occurring in particle transport rather than in the accelerators or in the spatial distribution of sources~\cite{Genolini2017prl}; 2) the AMS-02 measurements showed beyond doubt that protons, helium and heavier nuclei need to be injected at the sources with slightly different spectra, a finding that is at odds with the basic idea of diffusive shock acceleration. 3) The spectrum of positrons in CRs~\cite{AMS02.2013.posfraction,AMS02.2019.positrons} is very different from what one would expect as secondary products of hadronic interactions, thereby requiring additional sources of these antimatter particles~\cite{Hooper2009jcap,Serpico2012aph,Bykov2017ssrv,Manconi2020prd,Evoli2020arxiv}. This might also be the case, although less clearly, for antiprotons~\cite{Boudaud2020prr,Heisig2020prr,AMS02.2016.pbar}. 4) The measurement of beryllium~\cite{AMS02.2018.libeb} provided a firm confirmation that the CR residence time in the Galaxy requires extended magnetized halos, with size $\gtrsim 5$~kpc~\cite{Evoli2020prd,Weinrich2020aaa}. 

In most of these lines of investigation, the main limitation to our capability to extract physical information from the high precision data provided by AMS-02 comes from uncertainties in the cross sections for fragmentation of nuclei in spallation reactions~\cite{Webber1998apj,Silberberg1998apj,Mashnik2004adspr,Moskalenko2001icrc,Moskalenko2003icrc,Moskalenko2003apj,Tomassetti2017prd,Genolini2018prc}. It is probably worth stressing that this is also one of the main sources of systematic error in the measurement of the flux of heavy nuclei, due to the fragmentation of those nuclei inside the experiment, an effect that needs to be corrected for. 

The recent publication of the results of the measurement by AMS-02 of the spectra of intermediate mass nuclei in CRs~\cite{AMS02.2019.nemgsi} up to rigidities of order $\sim$TV, provides us with the most complete set of information about CR transport in that such nuclei are the parent particles for lighter nuclei through processes of spallation (for stable nuclei) and decay (for unstable isotopes). Such measurements extend now to iron nuclei, which have a predominantly primary nature~\cite{AMS02.2021.iron}. 

Provided particles are transported either diffusively or advectively over distances much larger than the thickness of the disc ($h\sim 150$ pc), the transport of CRs in the Galaxy is properly described by a weighted slab model~\cite{Jones2001apj,Aloisio2013jcap}, recently used to derive the basic properties of the CR transport (diffusion coefficient and advection velocity)~\cite{Evoli2019prd} and the size of the confining halo $H\gtrsim 5$ kpc, from the measurement of the Be/B ratio~\cite{Evoli2020prd}. 

This last goal could be achieved because of the fact that the $^{10}$Be unstable isotope decays to $^{10}$B, so that both numerator and denominator of the Be/B ratio are affected. The idea that this decay (and those of similarly unstable isotopes of other nuclei) could be used to infer the halo size dates back to~\cite{Ptuskin1998aap,Webber1998apj,Putze2010aap,Strong1998apj,Trotta2011apj} although such calculations were typically carried out using the $^{10}$Be/$^{9}$Be ratio (or similar ratios) at low energies. Notice also that some early papers used the Leaky Box model to derive the confinement time in the halo, a procedure that is known to lead to incorrect estimates. 

AMS-02 cannot separately measure the flux of these isotopes, but on the other hand it can measure the Be/B ratio up to high energies, $\sim$~TeV/n, where the size of the halo $H$ can be inferred without the huge uncertainties that derive from lack of information on CR transport in the disc, which is crucial for unstable nuclei at $\lesssim 100$ MeV/n energies.

The set of calculations discussed above showed that a consistent picture of CR transport can be obtained by analyzing the spectra of primary and secondary elements lighter than oxygen, while the secondary/primary ratios allowed us to determine the grammage traversed by CRs inside the Galaxy, thereby confirming that a break in the energy dependence of the diffusion coefficient is needed to describe AMS-02 data.

In fact the measurement of the beryllium flux and of the Be/B ratio led to a slight readjustment of the parameters due to small production of boron from the decay of the unstable $^{10}$Be isotope~\cite{Evoli2020prd} (this effect on the boron spectrum was noticed early on by~\cite{GarciaMunoz1987apjs}). As stressed above, the main limitation to the strength of these conclusions derives from the experimentally poor knowledge of the spallation cross sections, especially the partial cross sections for specific channels of fragmentation. A discussion of these uncertainties was provided in~\cite{Tomassetti2012apj,Genolini2018prc,Evoli2018jcap}. 
Another source of uncertainty, though to a smaller extent, was the fact that the abundances of heavier elements, acting as primaries for lighter nuclei, were poorly known and had to be taken from older experiments. This gap has been filled by the recent measurements of the fluxes of intermediate mass and heavy elements by AMS-02. 

In this article we apply the same approach, already mastered in~\cite{Evoli2019prd,Evoli2020prd}, to heavy nuclei, up to iron. We show that the spectra of all nuclei heavier than oxygen are described within uncertainties with an injection spectrum at the sources that is the same as for lighter intermediate mass nuclei (CNO), provided the contribution of spallation from even heavier nuclei is taken into account. This is in some tension with the statement by the AMS-02 collaboration that Ne, Mg and Si are three primaries that behave peculiarly, in that they have a different spectral index than He, C and O. We maintain that this behaviour is fully accounted for by properly calculating the secondary contribution to these nuclei from their heavier parent nuclei, a situation similar to that of nitrogen. 

The only exception to this general conclusion is the spectrum of iron as measured by AMS-02, which appears to be at odds with the predictions of the standard model. The main anomaly is a hard shape of the Fe spectrum for $R\lesssim 100$ GV, which is not compatible with the grammage inferred from secondary/primary ratios for lighter nuclei. Moreover, the ratio Fe/O measured by AMS-02 appears to be in tension with previous results from HEAO~\cite{Engelmann1990aa} and Voyager~\cite{Cummings2016apj}. We present here the results of several tests proposed to address these problems.

The article is structured as follows: in \S~\ref{sec:model} we provide a brief summary of the theoretical approach adopted here, based on the weighted slab technique applied to the whole chain of nuclei, from H to Fe. In \S~\ref{sec:method} we describe the methodology used to achieve a description of the nuclear spectra, as well as the secondary/primary ratios. In \S~\ref{sec:results} we describe in detail our findings, starting from nuclei of intermediate mass (Ne, Mg, Si) and going up to iron. We summarize our conclusions in \S~\ref{sec:conclusions}.

\section{Propagation model}
\label{sec:model}

The details of the transport equation and the approach we follow to solve it are detailed in~\cite{Jones2001apj,Aloisio2013jcap,Evoli2019prd,Evoli2020prd}. Here, we simply summarize the most important features of the model.

We assume that the CR density of stable and unstable nuclear species obeys a steady-state diffusion-advection equation~\cite{TheBible}:
\begin{multline}\label{eq:slab}
-\frac{\partial}{\partial z} \left[ D_{a} \frac{\partial f_a }{\partial z} \right]
+ v_A \frac{\partial f_a}{\partial z}
- \frac{dv_A}{dz} \frac{p}{3} \frac{\partial f_a}{\partial p}
\\
+ \frac{1}{p^{2}} \frac{\partial}{\partial p} \left[ p^{2} \left(\frac{dp}{dt}\right)_{a,\rm ion} f_a \right]
+ \frac{\mu v(p) \sigma_a}{m} \delta(z) f_a
+ \frac{f_a}{\hat\tau_{d,a}}
\\
= 2 h_d q_{0,a}(p) \delta(z)
+ \sum_{a' > a} \frac{\mu\, v(p) \sigma_{a' \to a}}{m}\delta(z) f_{a'}
+ \sum_{a' > a} \frac{f_{a'}}{\hat\tau_{d,a'}},
\end{multline}
where $f_a (p,z)$ is the distribution function of species $a$ in phase space, $v(p)=\beta(p) c$ is the particles' velocity, and $\mu = 2.3\,$mg/cm$^2$ is the surface density of the disk. 

The transport equation in Eq.~\ref{eq:slab} is obtained assuming spatially homogeneous diffusion and that the sources of CRs and the target gas are confined in an infinitely thin disc (half-thickness $h \simeq 100$~pc), in which all the interactions (spallation and ionization energy losses) occur. 

We assume that CRs are confined in a low-density infinite slab of half-thickness $H$, extending well beyond the gaseous disc ($H \gg h$). Outside the magnetic halo, the particles can escape freely into intergalactic space so as to reduce the CR density to $f_a(p, z = \pm H) \simeq 0$.

For the diffusion coefficient we adopt the parametrisation proposed in~\cite{Evoli2020prd}:
\begin{equation}\label{eq:diffusion}
D(R) = 2 v_A H + \beta D_0 \frac{(R/{\rm GV})^\delta}{[1+(R/R_b)^{\Delta \delta /s}]^s},
\end{equation}
where $D_0$ and $\delta$ are fitted to secondary over primary ratios, while the other parameters $s$, $\Delta \delta$ and $R_{b}$ are fixed from observations of primary nuclei. 
The change of slope above the rigidity $R_b$ in Eq.~\ref{eq:diffusion} has been implemented to account for the presence of similar breaks in the spectra of primaries~\cite{AMS02.2017.heco} and in the secondary/primary ratios~\cite{AMS02.2018.libeb}. As shown in~\cite{Genolini2017prl}, the spectral breaks seen in CR data at a rigidity of $\sim 300$~GV must be attributed to a change of properties in the Galactic transport rather than in the injection.
While a break in the diffusion coefficient was hypothesized by several authors, even before AMS-02 data \cite{Vladimirov2012apj,Evoli2012prd}, theoretical investigations have been carried out immediately after, to motivate this functional form in terms of the transition from self-generated turbulence to preexisting turbulence~\cite{Blasi2012prl,Aloisio2013jcap}, or of a non-separable spatially dependent diffusion coefficient in the Galactic halo~\cite{Tomassetti2012apj}. 
The plateau at low energies, where advection dominates transport, was found as a consequence of self-generated turbulence as described in~\cite{Recchia2016mnras}. Advection of CRs with self-generated waves is an unavoidable phenomenon peculiar of these models and virtually incompatible with the existence of second order Fermi acceleration on Galactic scales.

The advection velocity $v_A$ is assumed to be constant above and below the disk, hence adiabatic energy losses (third term in the LHS of Eq.~\ref{eq:slab}) are determined through the expression $dv_A/dz = 2v_{A} \delta(z)$.

The injection of CR nuclei of type $a$ is given as a power law in momentum, having in mind that they are predominantly accelerated at SN shocks~\cite{Blasi2013aarv,Morlino2017book,Blasi2019ncimr}, hence the source term can be written as $q_{0,a} \propto p^{-\gamma_{\rm inj}}$.
The slope $\gamma_{\rm inj}$ is assumed to be the same for all nuclei heavier than He, while the normalizations are chosen for each individual species to match the data. 
Notice that the assumption of universal injection for all nuclei, although strongly supported by theoretical grounds, does not apply to light CRs, as it has been shown that the injection of protons(helium) requires a softer(harder) slope than nuclei~\cite{Evoli2019prd,Weinrich2020aab}.

The other terms on the LHS of Eq.~\ref{eq:slab} describe the disappearance of the species $a$ because of spallation ($\sigma_a$) or of radioactive decay ($\hat\tau_{d,a}$). In the RHS, one can find the corresponding terms describing the production of the species $a$ by spallation and decay of heavier ($a' > a$) elements.
The inelastic spallation cross sections are computed using the parametric fits provided by~\cite{Tripathi1996nimpb,Tripathi1997nimpb}, while for the secondary production cross-sections we adopted the parametric fits of~\cite{Evoli2019prd} for the major primary channels of Li, Be and B production, and the approach described in~\cite{Evoli2018jcap} to compute all other cross-sections (including the cumulative contribution of ghost nuclei).
We remind that this approach is based on cross-section measurements (when available) to set the normalization of each spallation channel. 
To this aim, we make use of the GALPROP cross-section measurement database~\cite{Strong2001adspr,Moskalenko2001icrc,Moskalenko2003icrc} and a collection of more recent measurements~\cite{VillagrasaCanton2007prc}, in particular for the production of heavy fragments (Ne, Mg, Si, S) by iron spallation~\footnote{In order to facilitate the reproducibility of our results we release the cumulative cross-sections adopted to compute the secondary production via a dedicated repository~\cite{EvoliXsecs}}.

The quantities $\hat \tau_{d, a}=\gamma \tau_{d,a}$ define the Lorentz boosted decay lifetimes of unstable elements.
In~\cite{Evoli2020prd}, some of us addressed the effect of including the decay of $^{10}$Be, and the corresponding production of $^{10}$B, on the calculation of the Boron-over-Carbon ratio (B/C) and on the determination of the transport parameters.
Since the CR propagation in the environment of the Galactic disc might take place under different conditions, hostile to diffusion, we studied the minimum rigidity above which the $^{10}$Be decay takes place predominantly outside the disc. We found that this condition is well satisfied for rigidities $\gtrsim$ few GV's even for halos as small as $H \sim 1$~kpc \cite{Evoli2020prd}.
Above this rigidity, the Beryllium-over-Boron ratio as measured by AMS-02 was found to be best reproduced assuming $H\sim 7$ kpc, which we assume to be fixed here.

As in this work we focus on the propagation of heavy elements, additional unstable isotopes heavier than oxygen must be considered.
The unstable isotopes having lifetimes comparable with the propagation timescales at these energies are the $\beta^-$ unstable isotopes $^{26}$Al, $^{36}$Cl and $^{54}$Mn (see also~\cite{Donato2002aa}).

The shortest lifetime is the one of $^{36}$Cl, which is a factor $\sim 5$ smaller than the one of Be.
From Fig.~1 in~\cite{Evoli2020prd} we conclude that for halo sizes $H\gtrsim 6\,$kpc the condition that the nucleus decays outside the disc of the Galaxy is fulfilled above a few GV.

All other unstable isotopes have lifetimes at rest shorter than few kyrs and are treated as ghost nuclei, which means that we assume their instantaneous decay.

Finally, we comment on the fact that we do not model the decay modes involving electron capture, since it is important only at energies much smaller than $\sim$~GeV/n~\cite{Letaw1985apss}.

We account for the effect of solar modulation by using the force field approximation~\cite{Gleeson1968apj} with a Fisk potential $\phi$.

\begin{figure}
\centering
\includegraphics[width=\columnwidth]{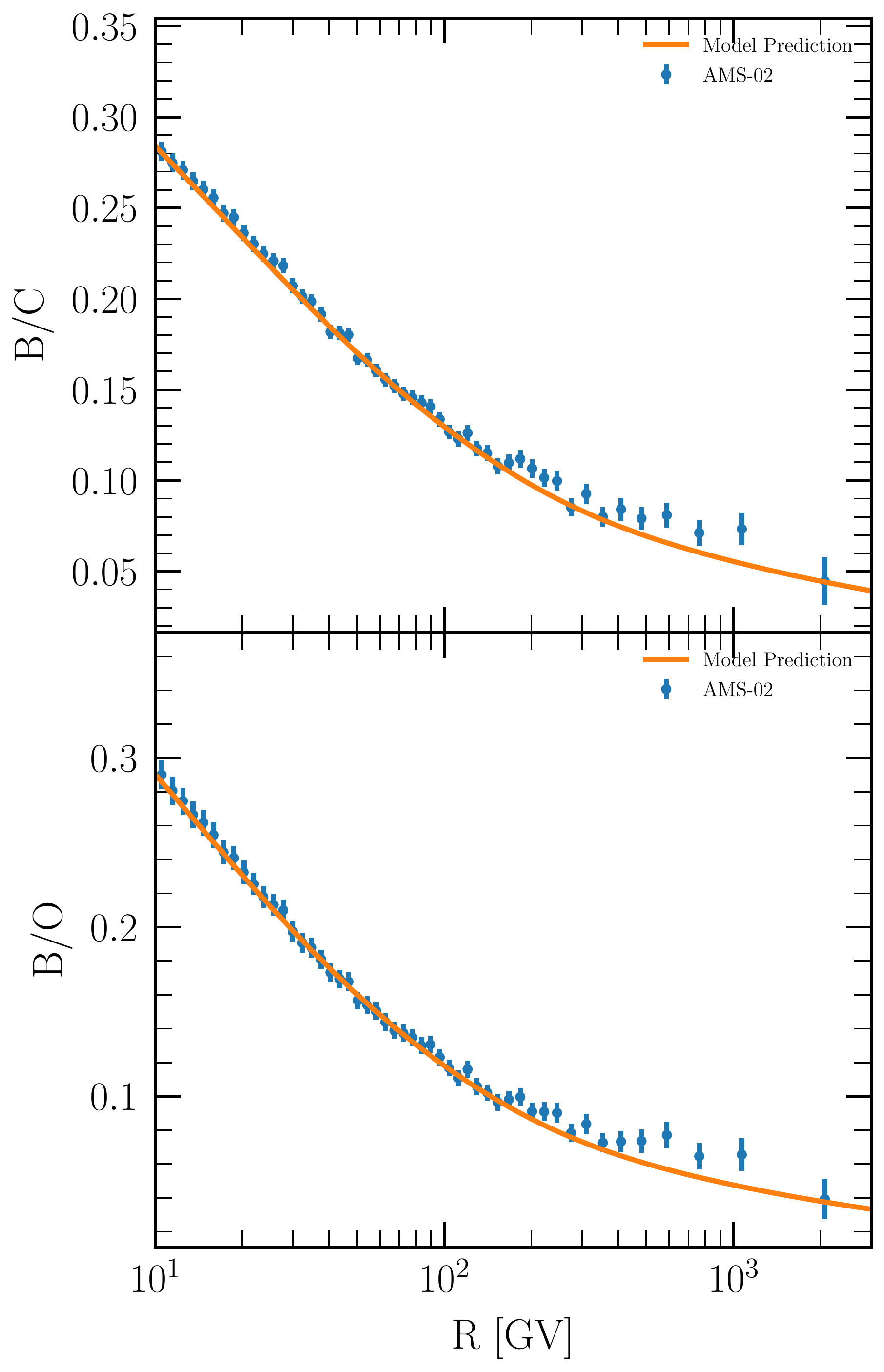}
\caption{B/C and B/O ratios as derived by using our calculations (solid line) compared to recent AMS-02 data~\cite{AMS02.2018.libeb}.}
\label{fig:light_ratios}
\end{figure}

\section{Methodology}\label{sec:method}

In order to get predictions for the fluxes of heavy elements, like Ne, Mg, Si, S and Fe, we include in our calculation all the stable isotopes from B ($Z=5$) to Ni ($Z=28$).	Additionally, we propagate the unstable isotopes $^{10}$Be, $^{26}$Al, $^{36}$Cl and $^{54}$Mn as described in \S~\ref{sec:model}. This procedure requires fitting of numerous parameters, that we summarize here. 
Spatial transport, including diffusion and advection, comprises 7 free-parameters: $D_0$, $\delta$, $v_A$, $H$, $R_b$, $\Delta \delta$, $s$. To speed up the analyses, as mentioned before, we fix the halo size to be consistent with the value derived from the analysis of the Be/B ratio, $H = 7$~kpc~\cite{Evoli2020prd}, having in mind that secondary-to-primary ratios are degenerate for $D_0/H$, to an excellent degree of approximation.
%
Moreover, we infer the three high-rigidity break parameters ($R_b$, $\Delta \delta$, $s$) with a fit to the H and He fluxes. Noticing that they are practically the same as the best-fit values of the recent AMS-02 publication \cite{AMS02.2021.review} , i.e. $s=0.09$ and $\Delta \delta=0.22$, obtained by fitting the proton spectrum.
However we require a somewhat lower $R_b$ of $290\,$GV to simultaneously fit He which is still compatible with the fit provided in \cite{AMS02.2021.review} within their error bars.
These assumptions reduce the number of free parameters which describe the transport to 3. 

The injection efficiencies $\epsilon_a$ of the species  C, N, O, Ne, Mg, Si, S and Fe are set in such a way that the total flux of each species match the AMS-02 data. That amounts to additional 8 parameters. For other primary species, in particular Na, Al, P, Ar, Ca, Cr, Mn and Ni, the TRACER03~\cite{Ave2008apj}, CRISIS~\cite{Young1981apj}, and HEAO3-C2~\cite{Engelmann1990aa} measurements have been used to fix their abundances and we do not include these normalizations as free parameters in the best-fit search. Notice that the efficiency $\epsilon_a$ pertains to the injection of a given charge, including all isotopes. We distribute the efficiency $\epsilon_a$ among the isotopes in order to reproduce the isotopic abundance in the interstellar medium (ISM) as measured in~\cite{LoddersPalme2009}.

The solar modulation potential, $\phi$, and the injection slope, $\gamma_{\rm inj}$ are two additional free parameters, bringing the total number of free parameters to 13. 

\begin{figure}
\centering
\includegraphics[width=\columnwidth]{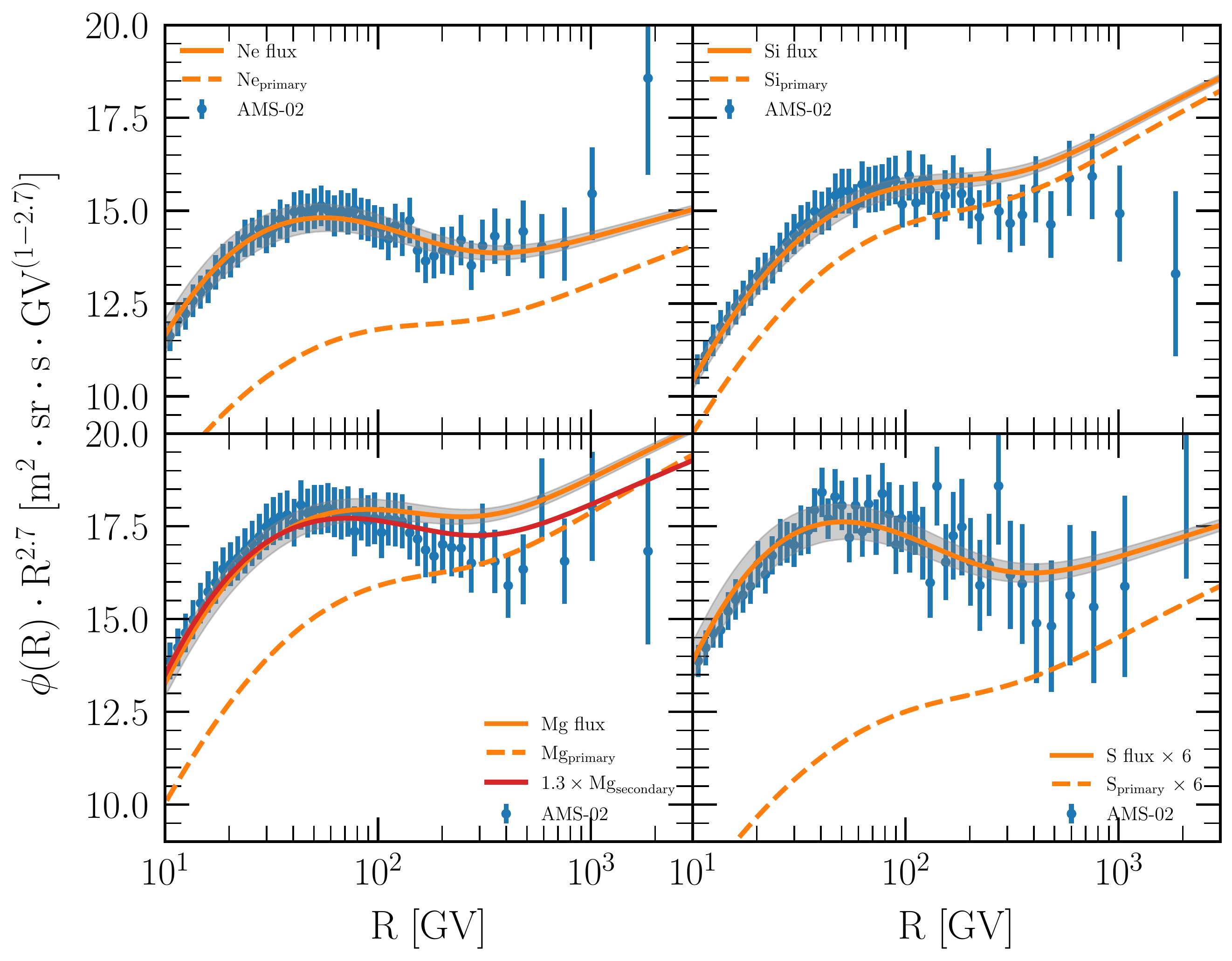}
\caption{Flux of Ne, Mg, Si and S (solid orange line) compared to recent AMS-02 data. The shaded area shows the effect of cross section uncertainties on the predictions. The dashed lines reflect the fluxes obtained by only including the primary contribution (no secondary production). The red line shows the expected flux for the given element, if the cross sections of the main production channels of Mg are increased by 30\% and the primary component of Mg is adjusted accordingly.}
\label{fig:fluxes}
\end{figure}

To fit the propagation parameters against different datasets we use the MINUIT package~\cite{JamesRoos1975}. To ensure that true minima are found, $O(50)$ minimisations from different starting points are carried out for all our analyses. The quantity we minimise is the $\chi^2$ computed for the AMS-02 measurements of different datasets. 
Specifically, the total $\chi^2$ is computed by summing the reduced $\chi^2$ computed over the ratios Be/C, B/C, Be/O, Ne/Mg, Si/Mg, Ne/O, Mg/O, Si/O, and the absolute fluxes of B, C, N, O, Ne, Mg, Si, and S~\cite{AMS02.2017.heco,AMS02.2018.libeb,AMS02.2019.nemgsi}. 
To normalize the S primary injection we make use of the AMS-02 preliminary data on the S absolute flux as reported in seminar presentations~\footnote{See, e.g., \url{https://indico.gssi.it/event/80/}}.
The absolute fluxes are fitted only for rigidities larger than $\sim 10$~GV in order to minimize the effect of solar modulation (see also~\cite{Evoli2019prd}).

Finally, for each dataset the total uncertainty is computed by adding systematic and statistic errors in quadrature~(see also~\cite{Weinrich2020aaa} for an attempt to take into account the correlation among systematics).

\section{Results}
\label{sec:results}

\subsection{Intermediate mass nuclei}
\label{sec:heavy}

\begin{figure}
\centering
\includegraphics[width=0.95\columnwidth]{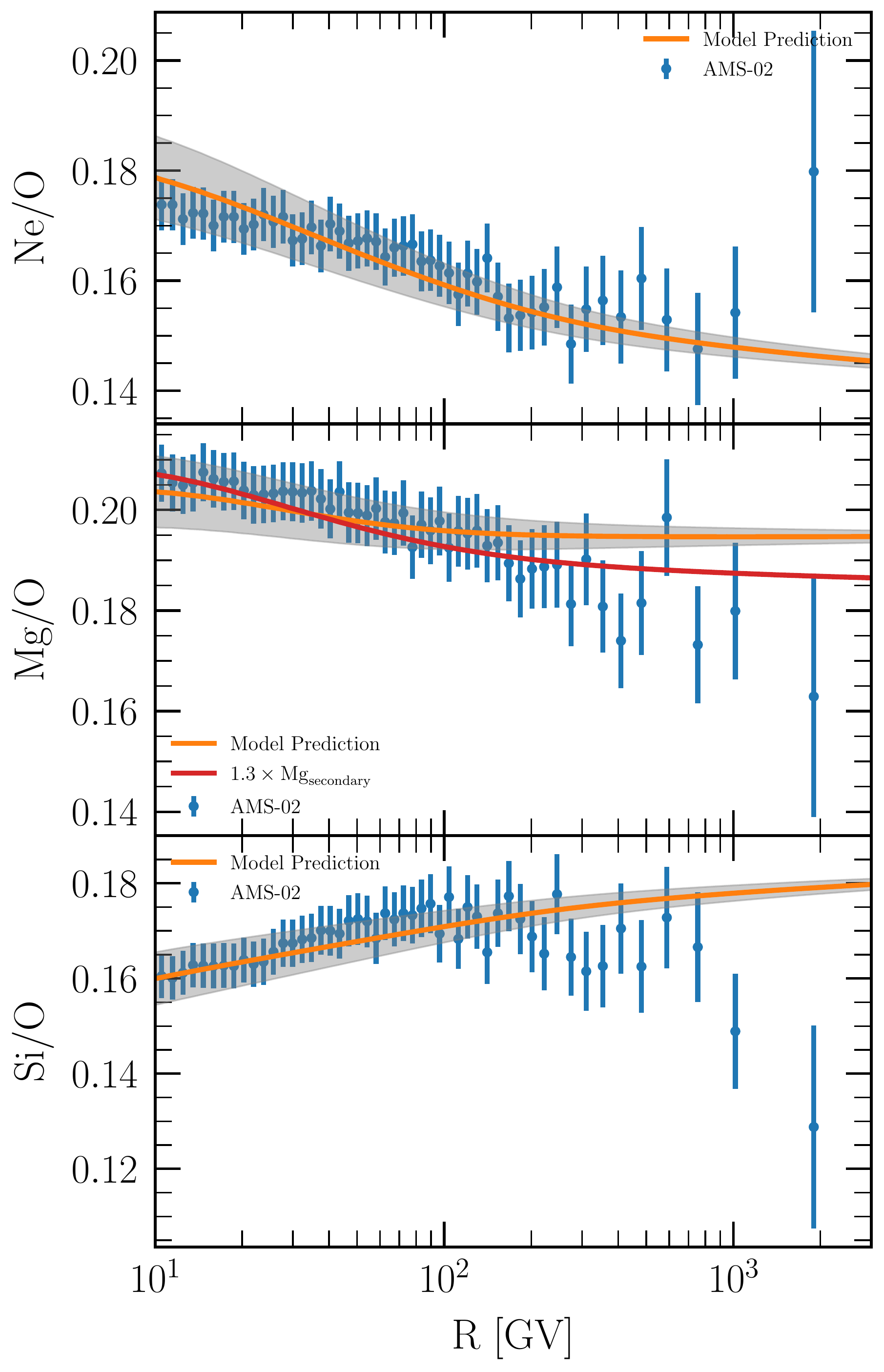}
\caption{Ratios of Ne, Si and Mg over O compared to recent AMS-02 data. The shaded region indicates the effect of cross section uncertainties on the model. The red line shows the case with higher Mg production cross sections.}
\label{fig:Oratios}
\end{figure}

The propagation parameters that best fit the data, using the method described in \S \ref{sec:method} are: $\delta = 0.56$, $\gamma_{\text{inj}} = 4.33$, $D_0/H = 0.35$ (in units of $10^{28}\,$cm$^2$ s$^{-1}$ kpc$^{-1}$, $v_A = 4.4\,$km/s and $\Phi = 0.49\,$GV. We notice that the best-fit values are in perfect agreement with the results in~\cite{Evoli2020prd} where the fit was performed using only the AMS-02 data on nuclei lighter than oxygen.

Following the procedure first introduced in~\cite{Evoli2019prd}, in order to illustrate the impact of cross section uncertainties on our results we have repeated our calculations 500 times for the best-fit scenario. 
In each repetition the spallation cross sections of elements heavier than oxygen are each rescaled by an individual factor. For each cross section the factor is sampled from a Gaussian distribution with mean $1$ and a width of $0.1$ for the total cross sections of a given element and $0.3$ for the partial cross sections of different channels respectively.
In all the following figures, the gray-shaded areas represent the 1-$\sigma$ variance of our calculations associated with the uncertainty in the cross-section.
In fact, the uncertainty in the cross sections might also affect the best-fit parameters, in particular the injection efficiencies, hence the error bands shown in our figures only capture part, though most likely the bigger part, of the effect of uncertainties in spallation cross sections.

In Fig.~\ref{fig:light_ratios} we show how the B/C and B/O ratios are correctly reproduced by our best-fit. The variance due to the uncertainty on the heavy-element cross sections is of the order of $\sim 1$\% at $\sim 10$~GV (and smaller at higher energies). This is due to the fact that the secondary contribution of elements heavier than oxygen to boron and carbon is sub-dominant compared to that of carbon, nitrogen, and oxygen~\cite{Evoli2018jcap}. This result justifies our previous attempts to extract the diffusion parameters by fitting the B/C and B/O even with a poor knowledge of the normalization of the flux of heavier elements.

The reduced $\chi^2$ for all the fits shown throughout this article is $\lesssim 1$, hence the overall description of data is excellent. On the other hand one might argue that a few data points in the B/C and B/O ratios at $R\gtrsim 400$ GV appear to be slightly displaced from the theoretical curve. This gives us the opportunity to point out that in this energy range there are at least three guaranteed effects that are expected to produce small deviations from the standard predictions (curves in Fig.~\ref{fig:light_ratios}): 1) CRs must experience some grammage inside the accelerators, which clearly depends on what the sources are.
For instance for the case of supernova remnants (SNRs) as sources of CRs, it was estimated in Ref. ~\cite{Aloisio2015aap} that the roughly energy-independent source grammage should be $\sim 0.1-0.2$~g/cm$^2$.
At $\sim 400$ GV this would cause a $\sim 5-10\%$ correction with respect to the standard estimate. 2) Some of the secondary nuclei produced around the acceleration region should be re-energized at the shock, which should somewhat change the estimate and energy dependence of the source grammage discussed above \cite{Berezhko2003aa,Blasi2009prl,BlasiSerpico2009prl}. 3) Secondary nuclei produced during Galactic transport occasionally encounter SNR shocks and get accelerated there. This effect, discussed in detail by \cite{Blasi2017mnras,Bresci2019mnras}, leads to small deviations from the standard expectation at high rigidities.

Although all these effects are guaranteed to play a role starting at $\sim$ few hundred GV rigidities, we do not have the capability to calculate them at the level of accuracy required to compare predictions with AMS-02 data. However, it is reassuring that even the standard calculations in the same rigidity range are uncertain at the $\sim 10\%$ level (Fig.~\ref{fig:light_ratios})).

Adopting the procedure discussed above, we obtain a best-fit that correctly reproduces the Ne, Mg, Si and the preliminary S data using the same injection slope as the intermediate nuclei, C and O. This can be appreciated in Fig.~\ref{fig:fluxes}. The case of Fe nuclei will be discussed later.
In each panel we show with a dashed line the result of our calculations obtained if the contribution to the flux of that nucleus from spallation of heavier elements were neglected. This allows us to immediately identify the species that are mainly primaries from those that receive a substantial secondary contribution: for instance, the Si flux is predominantly of primary origin (like C and O), while the Ne, Mg and S elements have mixed origin, resembling the case of Nitrogen. In all these cases the secondary production fills the low energy part of the spectrum, as one might easily have expected.

In Fig.~\ref{fig:fluxes}, a marginally worse agreement with data is seen in the case of Mg. However, given the uncertainties in the partial cross section for dominant channels of $\rm ^{26}Mg$ production from $\rm ^{28}Si$, we also studied the effect of increasing such cross sections by 30\% and reducing the source normalization correspondingly.
The resulting total Mg flux is shown as a solid red line in Fig.~\ref{fig:fluxes}, and agrees with AMS-02 data points at high energy visibly better than for the nominal value of the cross section for the branch $\rm^{28}Si\to {^{26}Mg}$.  

In Fig.~\ref{fig:Oratios} we show the ratios of Ne, Mg and Si over O, as functions of rigidity. The choice of plotting these ratios with respect to oxygen is due to the fact that oxygen can be considered (together with protons and iron) a pure primary species, meaning that the secondary contribution from spallation of heavier nuclei is negligible at all energies.
All of these ratios are in good agreement with measurements for $R\gtrsim 10\,$GV. 

\begin{figure}
\centering
\includegraphics[width=\columnwidth]{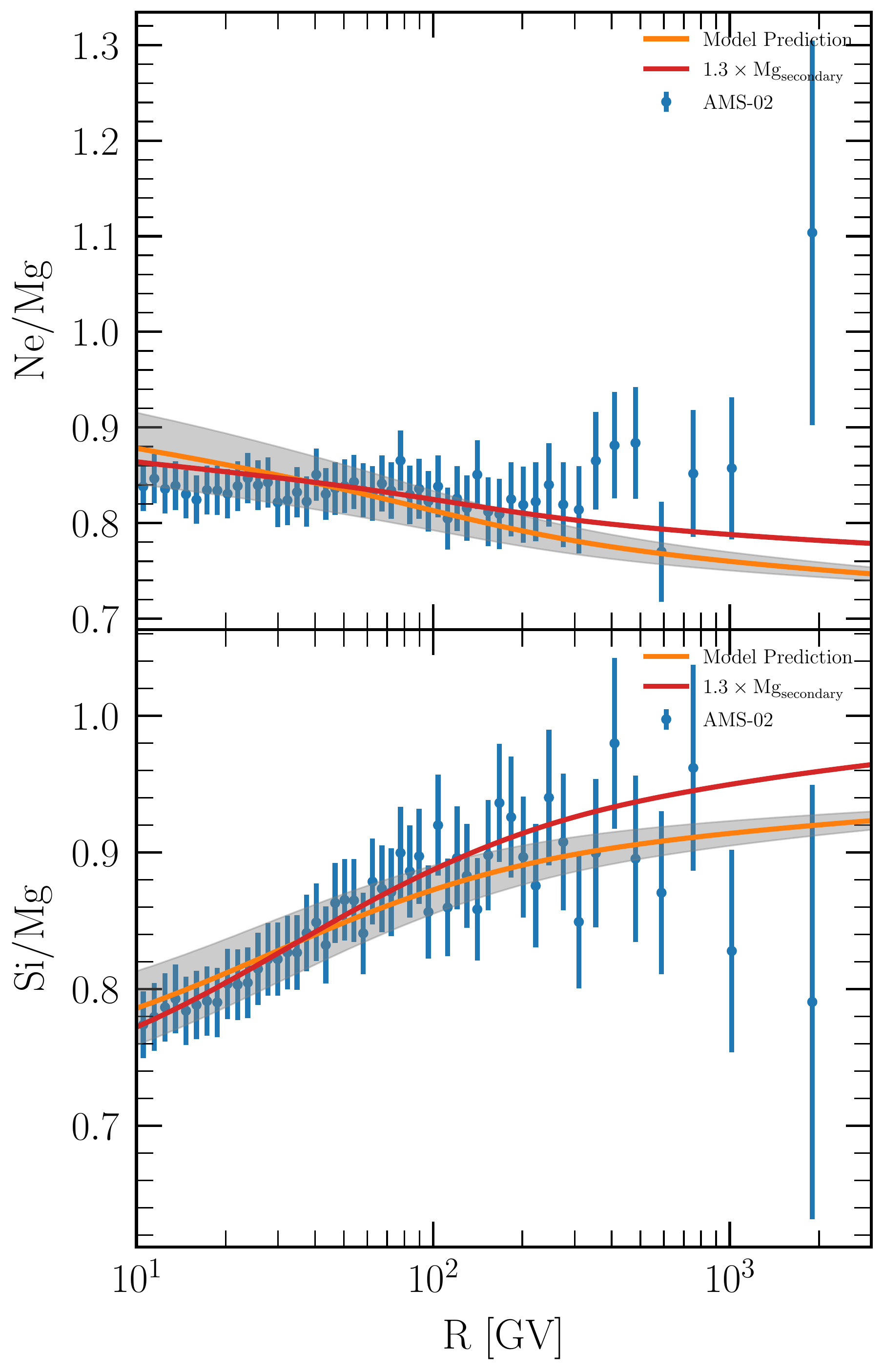}
\caption{Ratios of Ne and Si over Mg compared to recent AMS-02 data. The shadowed region indicates the effect of cross section uncertainties on the model. The red line shows the case with higher Mg production cross sections.}
\label{fig:ratios}
\end{figure}

Although Mg/O and Si/O slightly overshoot the data at very high rigidity, the experimental uncertainties of these data points are quite large, and we do not see this as evidence of a problem at this point. Moreover, for the case of the Mg/O ratio, we explored the effect of a possible increase by 30\% in the production cross section of Mg (solid red line, as in Fig.~\ref{fig:fluxes}): the increase of the cross section allows us to slightly reduce the normalization of the primary contribution to the flux of Mg, and hence to get a somewhat better agreement with data points. Similar considerations can be made for other heavy nuclei, for which the partial cross sections for spallation are known with comparably large uncertainty. In turn, this might be seen as a possible indication that our face values for at least some of these cross sections, might be underestimated. 

The case of Mg is a good illustration of the point already made earlier in this section: the shaded areas only catch part of the effect due to the uncertainties in the spallation partial cross sections. Changing these values leads to a change in the normalization of the primary contribution that is not accounted for in the shaded areas. 

In Fig.~\ref{fig:ratios} we also show the ratios of the Ne and Si fluxes over the Mg flux. Again, the agreement with data is remarkable. The only possible discrepancy with data could be identified in the Ne/Mg ratio at high rigidities. However, even this ratio is better described in terms of a mild increase in the production cross section of Mg, as discussed above (solid red line). 

In all these cases, it remains true that the main limitation to uncover the physics of CR transport and acceleration is in the poorly known cross sections for production of elements through spallation reactions. 

\subsection{Iron}
\label{sec:Fe}

Iron is as close to a primary nucleus as H and O are, hence the spectrum of iron can be considered as a reliable test of the injection spectra derived from the analysis illustrated above, and of the grammage traversed by CRs. The latter enters the calculation of the equilibrium spectrum of Fe through the severe spallation losses, which are expected to affect the flux of iron for $R\lesssim100$ GV, making it substantially harder than at higher rigidities. 

\begin{figure*}
\centering
\includegraphics[width=\columnwidth]{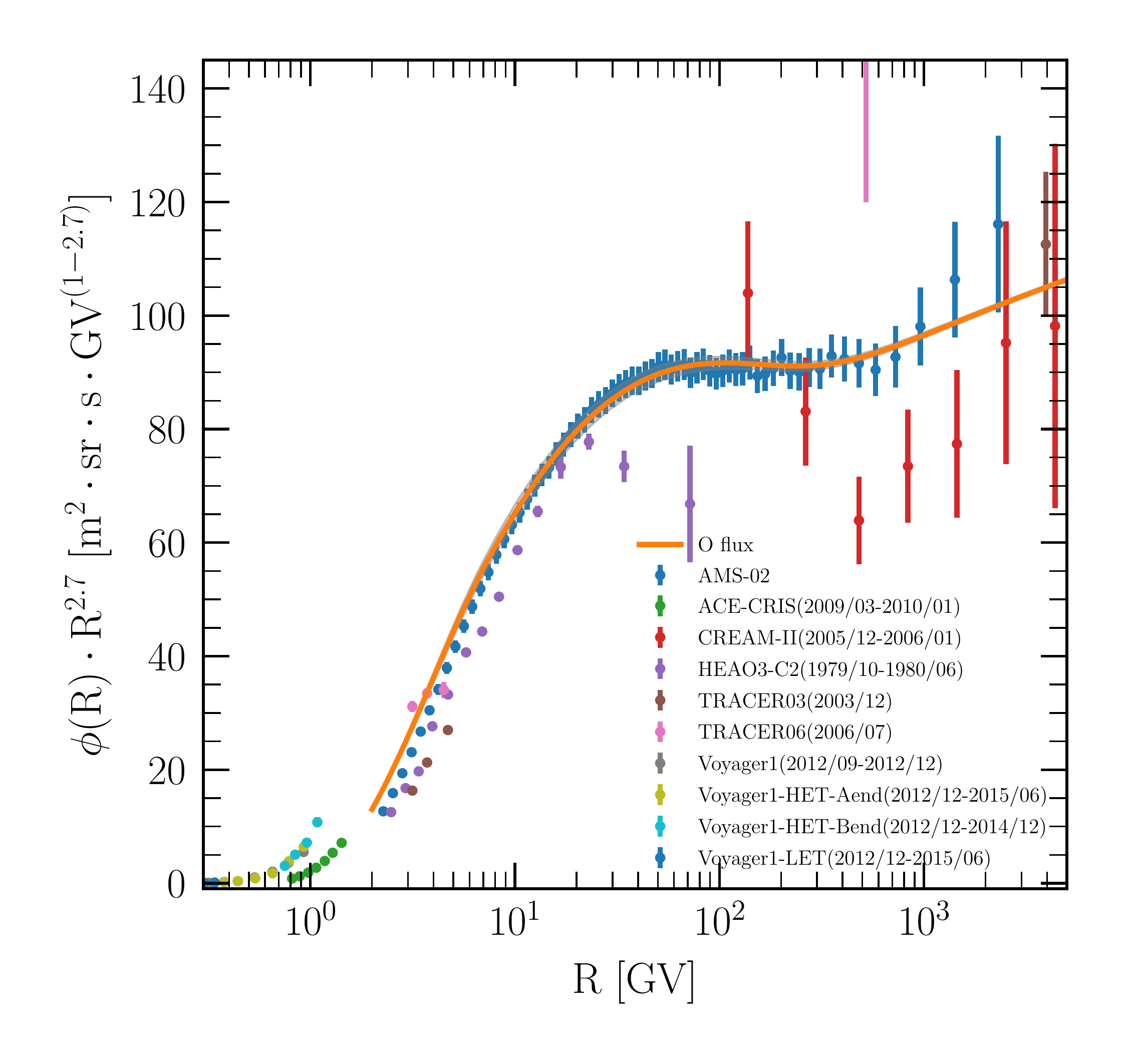}
\includegraphics[width=\columnwidth]{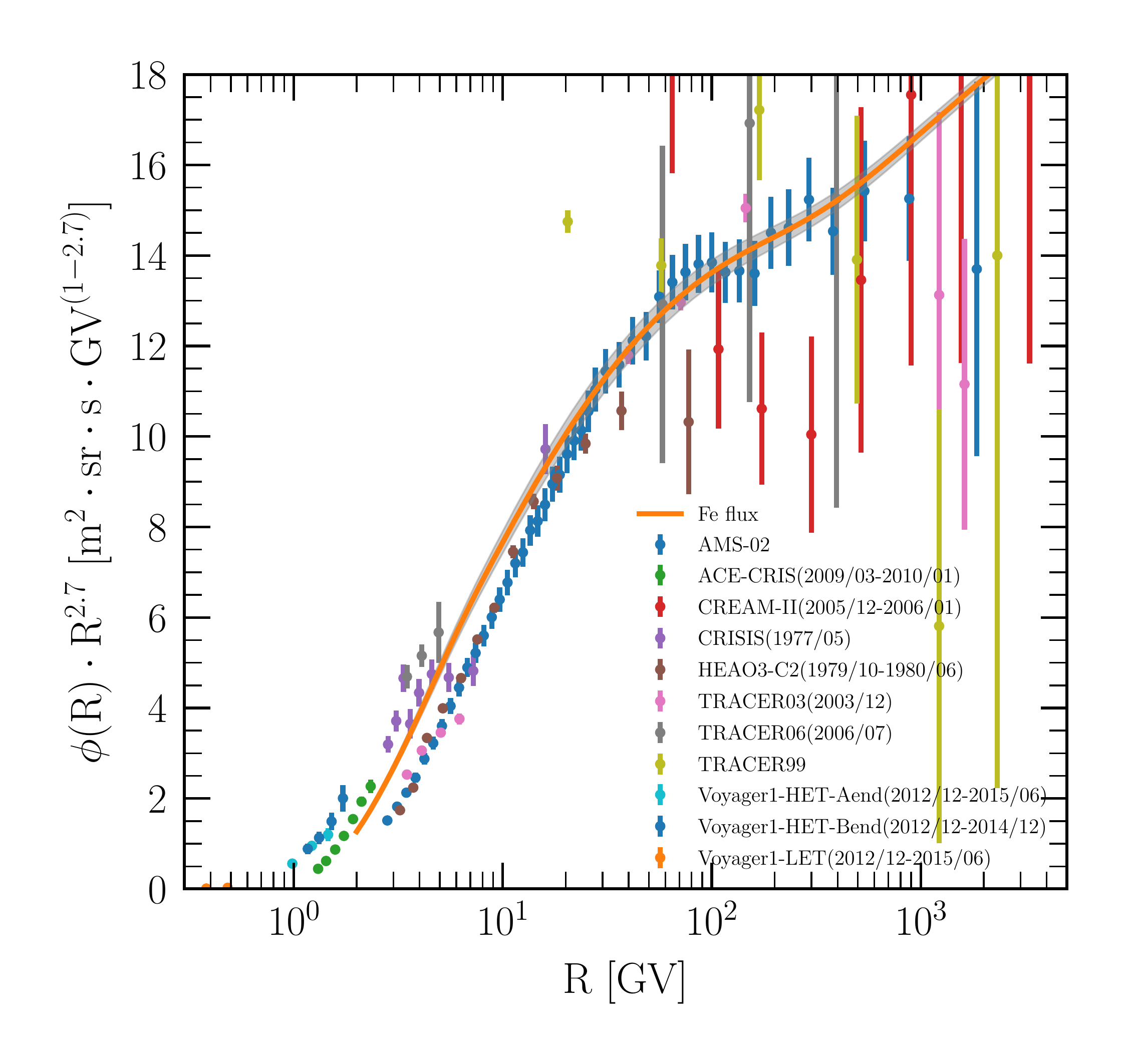}
\caption{Spectrum of oxygen (left) and iron nuclei (right) as predicted in our calculations (solid line), with an estimate of the uncertainties induced by the poor knowledge of cross sections (shaded area). The data points from AMS-02 and other experiments are also shown.}
\label{fig:OandFe}
\end{figure*}

The spectra of oxygen (left panel) and iron nuclei (right panel) that we derive using our calculations are shown in Fig.~\ref{fig:OandFe}, together with data points from AMS-02 and other experiments (see labels on the figure). A few comments are due: the spectrum of oxygen is very well reproduced at all energies, and the spectral break at $\sim 300$ GV is well visible, because the transport is dominated by diffusion in that rigidity range. For iron nuclei, the situation is different in that no clear evidence for a break is visible either in the data or in the prediction. This is expected since the transport of Fe nuclei is dominated by spallation even at $\sim 100$ GV. The comparison between predictions of our calculations and AMS-02 data clearly shows that for $R\lesssim 30$~GV, there is a strong disagreement, not reconcilable with error bars quoted by the ASM-02 collaboration. 

\begin{figure}
\centering
\includegraphics[width=\columnwidth]{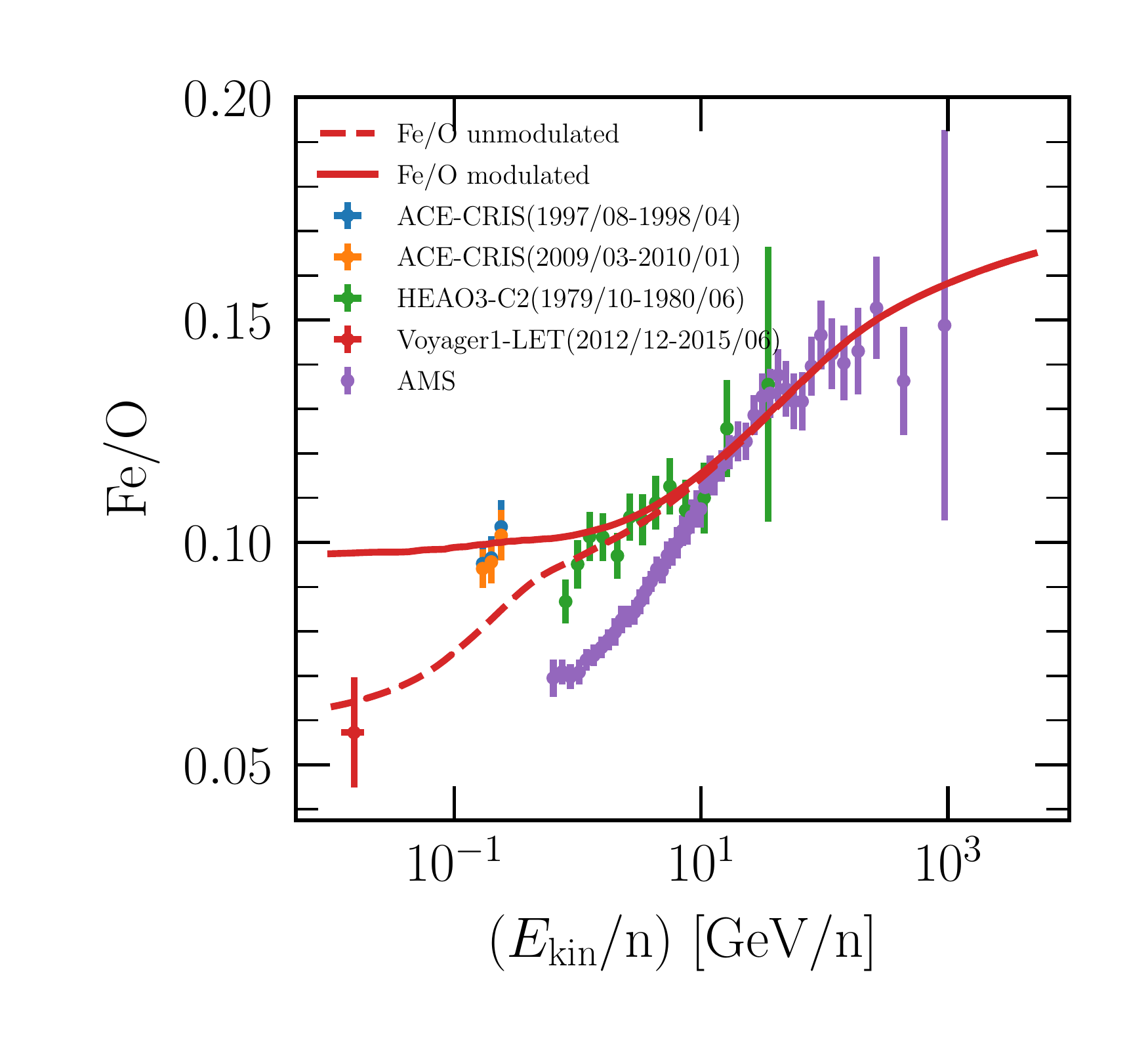}
\caption{Fe/O ratio of the modulated fluxes (solid line) as a function of kinetic energy per nucleon, compared with data from AMS-02, ACE-CRIS and HEAO. The dashed line shows the ratio of unmodulated fluxes, compared with the Voyager data.}
\label{fig:FeO}
\end{figure}

Following AMS-02~\cite{AMS02.2021.iron}, in Fig.~\ref{fig:FeO} we also show the Fe/O ratio as a function of kinetic energy per nucleon, which was also provided by the AMS-02 collaboration. In the same plot we show the same quantity as measured by ACE-CRIS~\cite{Lave2013apj}, HEAO3~\cite{Engelmann1990aa} and Voyager~\cite{Cummings2016apj}. The predicted ratio of modulated fluxes as derived using our calculations is shown as a solid (red) curve. Since the plot also contains a data point from Voyager, that measures unmodulated quantities, we also plot the results of our calculation of the Fe/O ratio of unmodulated fluxes (dashed line). 

The problem mentioned above at low energies is confirmed in the Fe/O ratio, which is predicted to be appreciably higher than AMS-02 measurements for $E_{kin}/n\lesssim 20$ GeV/n. The discrepancy is $\sim 30-40\%$ at $\sim 1$ GeV/n, much larger than the quoted AMS-02 systematic uncertainty in the same energy region. 

On the other hand it is worth noticing how the same ratio, measured by HEAO3 is also at odds with the AMS-02 result, while agreeing with our predicted trend and normalization. The Fe/O ratio as measured by ACE-CRIS at lower energies is also quite higher than an extrapolation of the AMS-02 data to lower energies. The case of Voyager data, at even lower energies confirms that the Fe/O ratio seems lower, although it is worth keeping in mind that Voyager measured unmodulated fluxes. On the other hand, our dashed curve (ratio of unmodulated fluxes) is perfectly compatible with the Voyager data point, despite all our fitting procedures were made only using data above $10$ GeV/n. The small bump in the dashed line reflects the energy dependence of spallation cross sections at low energies.  A warning is needed here: our calculations are increasingly less trustworthy when moving to lower and lower energies, when ionization energy losses force the whole iron propagation to take place in the disc, that in our calculations is assumed to be infinitely thin. In fact, all calculations of CR transport should be taken with much caution when this happens, because the diffusion coefficient of CRs in the disc is very likely different from the one assumed on Galactic scales, due to the severe ion-neutral damping that suppresses CR scattering in the disc~\cite{Kulsrud1971apjl}.

Given the oddity of the Fe and Fe/O situation, especially when compared with the impressive agreement of our predictions with all other data presented by AMS-02 on nuclei, we decided to explore some possible caveats of our approach that might help unveiling the source of the discrepancy: 

{\it 1)} the weighted slab model used in our work to describe particles' transport neglects the grammage accumulated by CRs in the halo. This is a good approximation as long as the density in the halo, $n_H$, is much smaller than $n_d(h/H)\sim 2\times 10^{-2}\rm cm^{-3}$. This appears to be a good approximation since the expected density in the halo is thought to satisfy this condition~\cite{Ferriere2001rvmp}. We actually calculated the solution of the transport equation of Fe nuclei in the presence of target gas in the halo and found that the low energy solution changes only at the percent level for gas density in the halo of order $10^{-3}\rm cm^{-3}$. Hence, it is unlikely that the discrepancy between predictions and AMS-02 data may be attributed to oversimplified CR transport in the halo. These effects might become of some importance only for halo gas density much larger than those suggested by observations.

{\it 2)} The uncertainties in the spallation cross sections are definitely a limiting factor in all these calculations. However most uncertain are the partial cross sections for fragmentation of a nucleus $A$ into a nucleus $A'$, while the total cross sections are somewhat better known. Nevertheless, for the sake of completeness, we tried to increase the spallation cross section of Fe by $40\%$, but even such drastic change turned out into a bad fit to the spectrum of Fe nuclei. Hence, the disagreement between theory and AMS-02 data below 100 GV cannot be due to an underestimation of the Fe spallation cross section.

{\it 3)} Despite the lack of physical support to the idea that iron nuclei may be injected with a different injection spectrum compared with other nuclei, we adopted an agnostic attitude and tried to find a best fit injection spectrum that could improve the fit (after all this is what is required for H and He). We find that adopting a harder injection spectrum (as required by the low energy Fe AMS-02 data), leads to only a slightly better fit at lower energies, while making the high energy fit worse. 

{\it 4)} Even more aesthetically unappealing is the possibility that the solar modulation may take place in a different way for iron and for oxygen. Imposing this unnatural solution, one finds that a good fit would require a Fe modulation about 70\% stronger than for oxygen.

In a recent article~\cite{Boschini2021arxiv}, the authors have noticed the same problem with Fe nuclei when comparing AMS-02 data with GALPROP predictions. It was suggested that in order to fit the AMS-02 iron spectrum it is necessary to introduce multiple breaks in the injection (source) spectrum. In fact the need for these multiple breaks is common to other nuclei as well, as found by~\cite{Boschini2020apjs}, quite at odds with the results illustrated in the present article, where the source spectra are pure power laws.

In~\cite{Boschini2021arxiv}, even at high energies, where all nuclei are expected to behave in the same way under the effect of diffusion, iron is assumed to be injected with a spectrum different from that of other nuclei. Moreover, in order to accommodate the obvious disagreement of the predicted flux with Voyager data, they suggest that the contribution from few local sources could become dominant. Clearly it is difficult to disprove such a model, especially because no modeling of such contribution was discussed. We notice however that the contribution of local sources at energies of few GeV/n seems unlikely. For an iron nucleus at that energy the distance traveled under the effect of losses is $d_l=(D \tau_l)^{1/2}\sim 1$ kpc (where $\tau_l$ is the energy loss timescale including ioinization and spallation, see also~\cite{Morlino2020prd}), too large to expect that something special may be happening. If the authors of~\cite{Boschini2021arxiv} were referring to a fluctuation due to accidentally local and recent sources, then this would in general lead to even larger fluctuations at higher energies where the global fits illustrated above seem to provide an excellent description of the data, with no need for local sources.

Finally, we find the assumption of multiple breaks in the source spectrum in~\cite{Boschini2021arxiv} not only lacking a solid justification on physical grounds, but also rather puzzling in that the breaks are different for different elements, thereby suggesting no clear trend in the yet to be found underlying physics. 

We stress once more that the AMS-02 data on the Fe/O flux are not only hard to reconcile with the results of existing calculations of CR transport on Galactic scales, but also and perhaps more important, with the results of previous experimental endeavours.


\subsection{H and He}
\label{sec:HeH}

\begin{figure}
\centering
\includegraphics[width=\columnwidth]{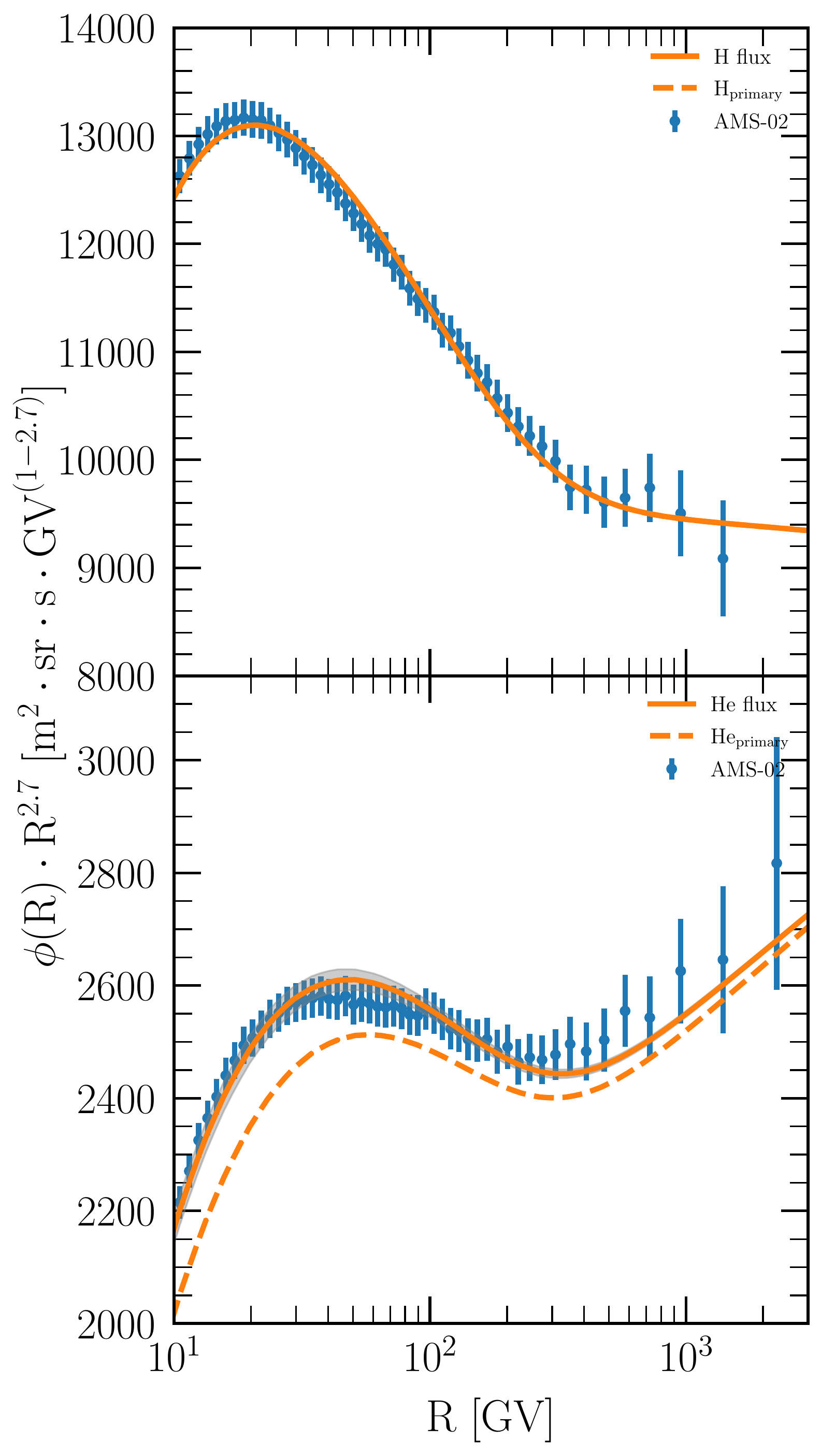}
\caption{Fluxes of H and He. The model parameters are the same of the last section but the injection slope of H is 4.37 and of He is 4.31. As in the other plots, the dashed line shows the predicted flux if the secondary production were neglected.}
\label{fig:HeH}
\end{figure}

\begin{figure}
\centering
\includegraphics[width=\columnwidth]{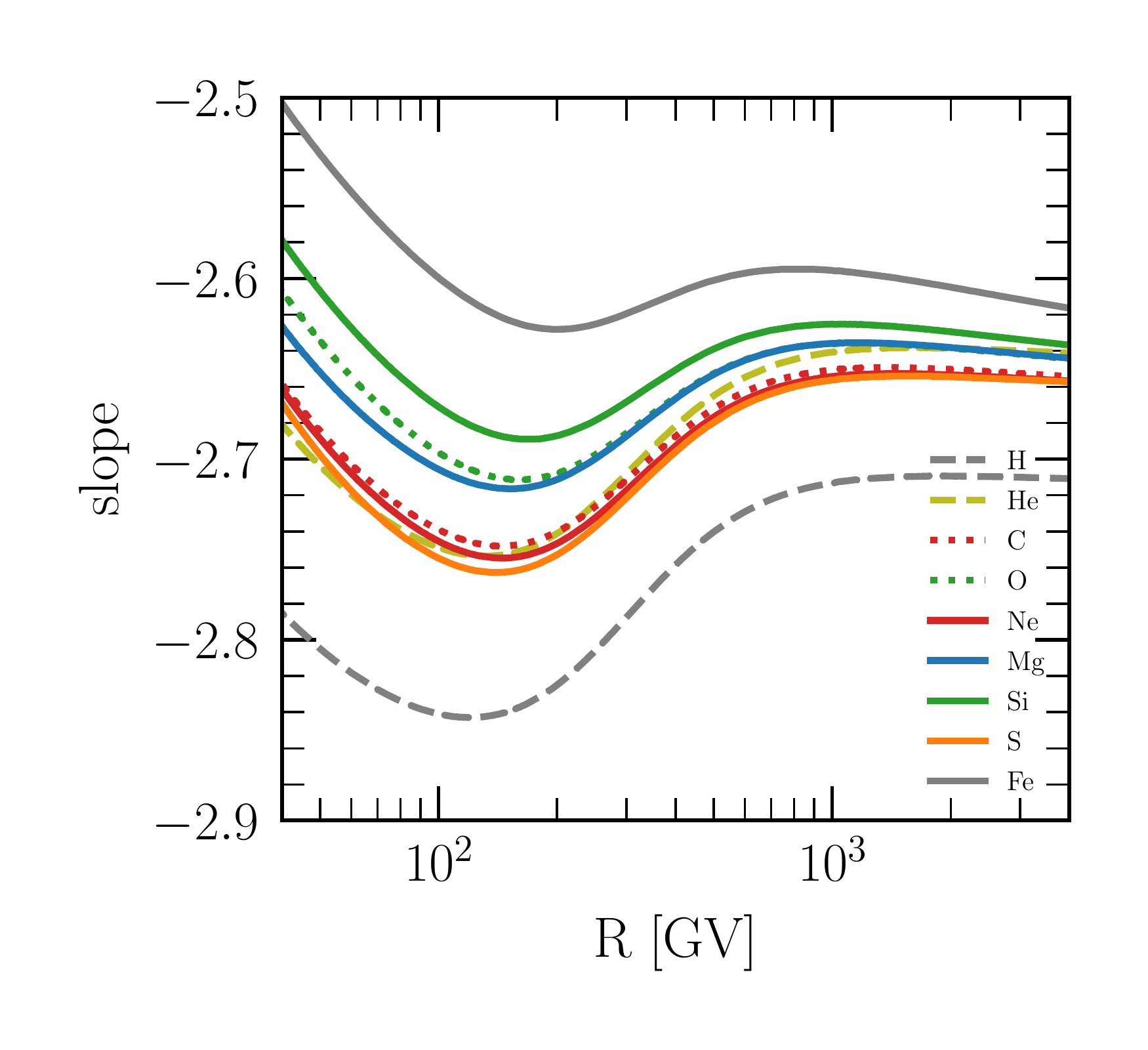}
\caption{Slope of the spectra of nuclei with different mass as a function of rigidity.}
\label{fig:slopes}
\end{figure}

As discussed in past literature, the situation of H and He is very peculiar and certainly puzzling. Here we confirm the conclusion of previous investigations ~\cite{Evoli2019prd} that AMS-02 data on H and He require different injection spectra. A similar conclusion was previously reached based on PAMELA data~\cite{PAMELA.2011.phe}. It turns out that H is best reproduced with a source spectrum with a slope which is $\approx 0.05$ softer than other nuclei, similar to what was found in~\cite{Evoli2019prd}, while He requires a source spectrum with a slope which is $\approx 0.02$ harder than that of nuclei.

At present no real explanation of this trend exists and certainly this result was not expected. Even models which suggest a different injection slope for the two cannot explain the difference in the injection spectra between He and intermediate mass nuclei, such as C and O~\cite{Malkov2012prl,Hanusch2019He}.

Notice that the remarkable similarity among the measured spectra of He, C and O nuclei (see Fig.~78 of~\cite{AMS02.2021.review}) is probably the most effective way to motivate the need for a different source spectrum, since the spallation of these three elements is quite different and yet the measured spectra are very similar.

Our best-fit to the H and He spectra is shown in Fig.~\ref{fig:HeH}, where, as usual, the shaded regions provide an estimate of the effect of uncertainties in the cross sections. The dashed line shows the primary contribution alone to the He flux, and its distance from the solid line shows the secondary contribution (take into account that the flux plotted in the figure is the sum of $^4\rm He$ and $^3\rm He$, and the latter is also produced in spallation reactions of $^4\rm He$). 

\section{Conclusions}
\label{sec:conclusions}

The AMS-02 data provide us with the tools for what could be the most substantial step forward in our understanding of the origin of CRs. In the present article we discussed in detail the implications of these data in terms of CR transport in the Galaxy. 

The general framework that emerges from our calculations is one in which three slopes are required to describe the source spectra for elements of different masses: a steep injection for protons, one sightly harder spectrum for nuclei heavier than He and one even harder for He itself. The difference between the slopes is at the level of few percents. This might seem a small difference, but the high precision data now available force us to provide a description of physical processes at this level. Taken at face value, this difference, however small, imposes a big strain on our models of the origin of CRs: all the acceleration processes that we have devised through the years are rigidity dependent, namely nuclei of the same rigidity should experience the same acceleration and manifest the same spectrum. The situation is somewhat more complex, in that the spectrum of particles released into the ISM is not the same as at the acceleration site (see for instance~\cite{Caprioli2009mnras,Caprioli2010aph,Cristofari2020aph}). But it remains true that nuclei of different mass but the same rigidity should escape the acceleration region in the same way. 

It has also been speculated that the integration over time during a SNR evolution might give rise to a difference in the spectra of H and He~\cite{Malkov2012prl,Hanusch2019He}. However, aside from the details of the model, it predicts that He and heavier nuclei should share the same spectrum.

It may also be speculated that different types of sources may release particles with slightly different spectra and different relative abundances (for instance some sources may be He-rich), but such solutions all appear very fuzzy at the present time, and it should be admitted that no clear understanding of the origin of the different source spectra for H, He and heavier nuclei is yet available. 

Once it is accepted that three injection slopes are necessary to describe observations, we have been able to show that the spectra of nuclei are all properly accounted for, with the possible exception of iron, that will be discussed later. In order to reach such a conclusion it is of the utmost importance that all chains of spallation reactions and radioactive decays are taken into account~\cite{Webber1998apj,Silberberg1998apj,Mashnik2004adspr,Moskalenko2001icrc,Moskalenko2003icrc,Moskalenko2003apj,Tomassetti2017prd,Genolini2018prc}. The cross sections for these processes, or rather the poorness with which such cross sections are currently known, represent the chief limitation to our capability to extract physical information from available data.  For some nuclei, for instance Mg, the comparison between observed and predicted spectra would benefit from slight increases in the production cross sections, compatible with known uncertainties.

The spectra that result from propagation also have a trend in terms of observed spectral shape, as illustrated in Fig.~\ref{fig:slopes}: 1) for increasingly more massive nuclei, there is a trend towards globally harder spectra, as a result of the enhanced role of spallation. This is particularly evident for nuclei which are mainly of primary origin (O, Si and Fe), while deviations from this trend can be noticed for nuclei which receive a substantial secondary contribution. 2) At low energies, where spallation reactions play a more important role the slope decreases for heavier nuclei. 3) For rigidities $R\gg 10^3$ GV all nuclei heavier than He tend to have the same spectral slope, since at such rigidities CR transport is dominated by diffusion, which behaves in the same way for all nuclei at given rigidity. Small differences around these trends can be easily identified (and have been discussed above) for nuclei that are substantially contaminated by the contribution of spallation from heavier nuclei (for instance this is the case of N~\cite{AMS02.2018.n}). For such nuclei, deviations are visible at low rigidities where the secondary contribution becomes important. 

The global picture of CR transport that emerges from our calculations is that, aside from energy losses, the phenomenon is well described by diffusion and advection. A few comments on both ingredients may be useful: the character of particle diffusion is fully reconstructed using the proton spectrum and secondary/primary ratios, in addition to the Be/B ratio that provides an estimate of the halo size~\cite{Evoli2019prd,Weinrich2020aab}. 
Both these pieces of observations suggest that the spectral breaks observed in spectra at $\sim 300$~GV are due to a corresponding break in the diffusion coefficient. Such a break was proposed to originate from a transition from self-generated to pre-existing turbulence in the Galaxy~\cite{Blasi2012prl,Aloisio2013jcap,Aloisio2015aap,Evoli2018prl} (see also \cite{Blasi2019galax,Amato2018adspr} for recent reviews) or from a non trivial spatial dependence of the diffusion coefficient \cite{Tomassetti2012apj}. At rigidity $R\lesssim 10$ GV, advection with Alfv\'en waves becomes important. Although in this article we do not include self-generation explicitly, it was previously shown~\cite{Blasi2012prl} that such terms cannot be eliminated in theories where CRs are allowed to generate their own scattering centers. Such models are also incompatible with having second order Fermi acceleration at low energies, since waves move preferentially away from the galactic disc.   

As mentioned above, the only exception to this otherwise satisfactory picture of CR transport is represented by the spectrum of iron nuclei measured by AMS-02~\cite{AMS02.2021.iron,Boschini2021arxiv} at $R\lesssim 30$ GV. In this region the discrepancy between our predictions and the measured spectra are at the level of $\sim 20-40\%$, much larger than the quoted systematic error in the same energy region. The discrepancy is emphasized even better in the Fe/O ratio, where AMS-02 data also seem to be at odds with HEAO data at the same kinetic energy and with ACE-CRIS, if extrapolated at lower energies. We explored several possible avenues to reduce the discrepancy between our predictions and AMS-02 data but none of them came close to be either a resolution of the problem or exemplary in plausibility. 

To add to this puzzling situation, the predictions based on our calculations smoothly connect the high energy Fe/O ratio to the ACE-CRIS data. Moreover, when the ratio Fe/O is calculated on unmodulated fluxes, as it would be appropriate for Voyager data, our predicted ratio is in impressive agreement with Voyager data points. 

At this point it is hard to draw any firm conclusion on the origin of this discrepancy, but there is a serious possibility that the problem with the iron spectrum may be of experimental nature and that only a detailed comparison between different techniques and different analyses can unveil the source of the disagreement. It is probably useful to stress that other discrepancies would deserve the same attention, such as the marked difference between the C and O spectra of AMS-02 compared with those of PAMELA~\cite{PAMELA.2014.bc} and CALET~\cite{CALET.2020.co}. At low energies the effects of nuclear fragmentation inside the experiment itself are rather large and for heavy elements, such as iron, this effect requires a careful correction. In this sense, it would be interesting and instructive to see a detailed description of the analysis of these data for different experiments, perhaps with different models for the relevant cross sections.

Several CR detectors, based on various detection techniques, are currently in operation and are expected to release the spectrum of O and Fe nuclei in the near future. These data will serve as additional tests of the nature of the discrepancy. 

\section*{Acknowledgements}
We are grateful to the anonymous referee for several very interesting comments that helped us improve the quality of the manuscript. We are grateful to R. Aloisio, M. Boezio, I. De Mitri and R. Munini for useful conversations. The research activity of PB was partially funded through Grant ASI/INAF n. 2017-14-H.O. We acknowledge the use of the CRDB~\cite{Maurin2020crdb} and ASI~\cite{DiFelice2017asi} databases of  CR measurements.

\end{document}